\documentclass[preprint,number]{elsarticle}

\usepackage[latin1]{inputenc}
\usepackage[english]{babel}
\usepackage{fancyhdr}
\usepackage{amsfonts}
\usepackage{bbm}
\usepackage{pdflscape}
\usepackage{amssymb,amsmath,mathrsfs,centernot}								
\usepackage{multicol}									
\usepackage{graphicx}									
\usepackage{enumerate}								
\usepackage{xcolor}										
\usepackage[bottom]{footmisc} 				
\usepackage{hyperref}									
\usepackage[nottoc,numbib]{tocbibind}
\usepackage{floatrow}
\usepackage{verbatim}
\usepackage{rotating}
\usepackage[labelfont={footnotesize,bf},textfont={footnotesize}]{caption}
 \usepackage[toc,page]{appendix}
  \usepackage{caption}
\usepackage{subcaption}
\usepackage{amsthm}

\hypersetup{colorlinks=true, citecolor=blue, linkcolor=blue, urlcolor=blue}	
\makeatletter
\def\ps@pprintTitle{%
  \let\@oddhead\@empty
  \let\@evenhead\@empty
  \let\@oddfoot\@empty
  \let\@evenfoot\@oddfoot
}
\makeatother

\newtheorem{theo}{Theorem}[section]

\begin{document}

\begin{frontmatter}

\title{Existence and Uniqueness of a Steady State for an OTC  Market with Several Assets}

\author[udes]{Alain B\'{e}langer\corref{ab}}
\author[udes]{Ndoun\'{e} Ndoun\'{e}}
\ead{ndoune.ndoune@usherbrooke.ca}

\cortext[ab]{Corresponding author: alain.a.belanger@usherbrooke.ca, tel. 819-821-8000 x62368, fax. 819-821-7934}
\address[udes]{Facult\'{e} d'administration, Universit\'{e} de Sherbrooke, Qu\'{e}bec, J1K2R1, Canada}

\date{\today}



\begin{abstract}
We introduce and study a class of over-the-counter market models specified by systems of Ordinary Differential Equations (ODE's), in the spirit of Duffie-G\^{a}rleanu-Pedersen \cite{Duffie2005}. The key innovation is allowing for multiple assets. We show the existence and uniqueness of a steady state for these ODE's.

\end{abstract}


\end{frontmatter}


\section{Introduction}

This article addresses the question of the existence and uniqueness of a steady state for Over-the-Counter (OTC) markets with any number of traded assets. The type of markets we study here are inspired by the widely cited and pioneering work of Duffie, G\^arleanu and Pedersen (see \cite{Duffie2005} and \cite{Duffie2007}).  Darrell Duffie's recent monograph, \textit{Dark Markets} (see \cite{Duffie2012}), documents some of the modelling efforts done to understand the dynamics of OTC markets.  It is an active area of research but Duffie also notes that it is still underdeveloped in contrast with the vast literature available on central market mechanisms.

 Our goal is to shed some light on foundational issues in asset pricing in OTC markets with several assets. In particular, we study models of OTC markets described by ODE's which have not yet appeared
 in the differential equations literature. It is well known that in OTC markets, an
investor who wishes to sell must search for a buyer, incurring opportunity and other
costs until one is found (see for instance \cite{Duffie2005}). For the case of one asset, the evolution of an investor's state can be described  by a system of four quadratic differential equations, an overview is given in Chapter 4 of \cite{Duffie2012}. In their original paper \cite{Duffie2005}, the authors computed explicitly the steady state of their system and they showed directly its uniqueness.  Here we study the more general case with several assets for an extended model which is still described by systems of quadratic differential equations.

In B\'elanger et al. \cite{Blanger2013}, another extension of the DGP model with several assets was considered and the existence and uniqueness of the steady state was established using the Intermediate Value Theorem. The authors  also derived the value functions of the investors in order to obtain the prices at which investors trade with each other at that steady state and they showed moreover that the steady state is (exponentially) stable.This simpler extension of the DGP model can be called non-segmented market since it does not track the asset an investor wants, when she enters the market. The extension was first considered by Weill in \cite{Weill2008} where the author studies the determinants of liquidity premia in dynamic bargaining markets.

The extension considered in this paper, which, following Vayanos-Wang \cite{Vayanos2007}, we call the partially segmented market with several assets, does keep track of the asset desired by an investor upon entering the market. A variant of this model was first considered by Vayanos-Wang \cite{Vayanos2007} who studied  liquidity premia in a search and bargaining market with two assets.

The OTC market model is described in section 2. In section 3, we show how to use  the generalized nonlinear alternative of Leray-Schauder-Krasnosel'skij \cite{Djebali2010} to obtain the existence of a steady state. Finally, we establish the uniquenes of the state in section 4  using Kellogg's uniqueness theorem \cite{Kellogg1976}. The traded prices at this steady state and its asymptotic stability will be addressed in future work.

\section{Description of the  model}

In their seminal paper of 2005, Duffie et al. \cite{Duffie2005} present their model of OTC market with one traded asset  as a system of four linear ODE's with two constraints which can be
reduced to a system of two differential equations
with two constraints. In this section, we describe an extension of their model
involving $K$ traded assets, $K\geq 1$.

The set of available assets will be denoted $\mathcal{I} = \{1,...,K\}$. Investors can hold at most one unit of any asset $i\in\mathcal{I}$ and cannot short-sell. Time is treated continuously and runs forever. The market is populated by a continuum of investors. At each time, an investor is characterized by whether he owns the $i$-th asset or not, and by an intrinsic type which is either a 'high' or a 'low' liquidity state. Our interpretation of liquidity state is the same as in \cite{Duffie2005}. For example, a low-type investor who owns an asset may have a need for cash and thus wants to liquidate his position. A high-type investor who does not own an asset may want to buy the asset if he has enough cash. Through time, investors' ownerships will switch randomly because of meetings leading to trades and the investor's intrinsic type will change independently  via an autonomous movement. This dynamics of investor's type change is modeled by a (non-homogeneous) continuous-time Markov chain $Z(t)$ on the finite set of states $E$. The  state of  an investor is given by an element of $E = \{(l,n),(hi,o),(hi,n),(li,o)\}_{i\in\mathcal{I}}$, where the first letter designates the investor's intrinsic liquidity state and the second designates whether the investor owns the asset $i$ or not. As noted in the introduction and in contrast to the non-segmented market model of
\cite{Begm2013}, buyers in this model enter the market with the intention of purchasing a specific asset $i$. If an investor initially does not own any asset and is a low-type, the switching intensity of becoming a high-type is denoted $\widetilde{\gamma}_{ui}$ and it now depends on the asset type. If he initially does not own any asset but is a high-type, his switching intensity of becoming a low-type is denoted by $\widetilde{\gamma}_{di}$ and it now also depends on the asset. However, if an investor initially owns the specific asset $i$ and is a high-type, the switching intensity of becoming a low-type is denoted by $\gamma_{di}$. Finally, if he initially owns a specific asset $i$ but is a low-type, the switching intensity of becoming a high-type is $\gamma_{ui}$. We make the liquidity switches depend on the asset because these assets could have different  purchase prices and  dividend flows.
In addition, investors meet each other at rate $\lambda_i$, but an exchange of the asset occurs only if an investor of type $(li,o)$ meets one of type $(hi,n)$.
One should notice that, without changes of positions, the system would stop after a finite time and the market would become inefficient.
 At any given time $t$, let $\mu_t(z)$ denote the proportion of investors in state $z \in E$. So, for each $t \geq 0$, $\mu_t$ is a probability law on $E$.

Let $m_i$ denote the proportion of asset $i$, for all $i \in \mathcal{I}$.

 We can now describe  the dynamical system of investors' type proportion measure $\mu_t(z)$ for each $z\in E$, which consists of $3K+1$ equations:
\begin{eqnarray}
	\dot{\mu}_t(hi,n) &=& -\lambda_i \mu_t(hi,n)\mu_t(li,o) + \widetilde{\gamma}_{ui} \mu_t(l,n) - \widetilde{\gamma}_{di}\mu_t(hi,n), \ \forall i\in\mathcal{I}\label{peq1}\\
	\dot{\mu}_t(l,n) &=&  \sum_{i\in\mathcal{I}} \lambda_i \mu_t(hi,n) \mu_t(li,o)- \sum_{i\in\mathcal{I}}\widetilde{\gamma}_{ui} \mu_t(l,n) + \sum_{i\in\mathcal{I}}\widetilde{\gamma}_{di}\mu_t(hi,n)\label{peq2}\\
	\dot{\mu}_t(hi,o) &=& \lambda_i \mu_t(hi,n)\mu_t(li,o) + \gamma_{ui}\mu_t(li,o) - \gamma_{di}\mu_t(hi,o), \ \forall i\in\mathcal{I}\label{peq3}\\
	\dot{\mu}_t(li,o) &=& -\lambda_i \mu_t(hi,n)\mu_t(li,o) - \gamma_{ui} \mu_t(li,o) + \gamma_{di}\mu_t(hi,o), \ \forall i\in\mathcal{I}\label{peq4}
\end{eqnarray}
with the $K+1$ constraints
\begin{align*}
	\mu_t(hi,o) + \mu_t(li,o) &= m_i, \ \forall i\in\mathcal{I}\\
	\sum_{i\in\mathcal{I}}m_i  + \sum_{i\in\mathcal{I}} \mu_t(hi,n)  + \mu_t(l,n) &= 1
\end{align*}

\begin{figure}[h!]
  \centering
  \includegraphics[width=4in]{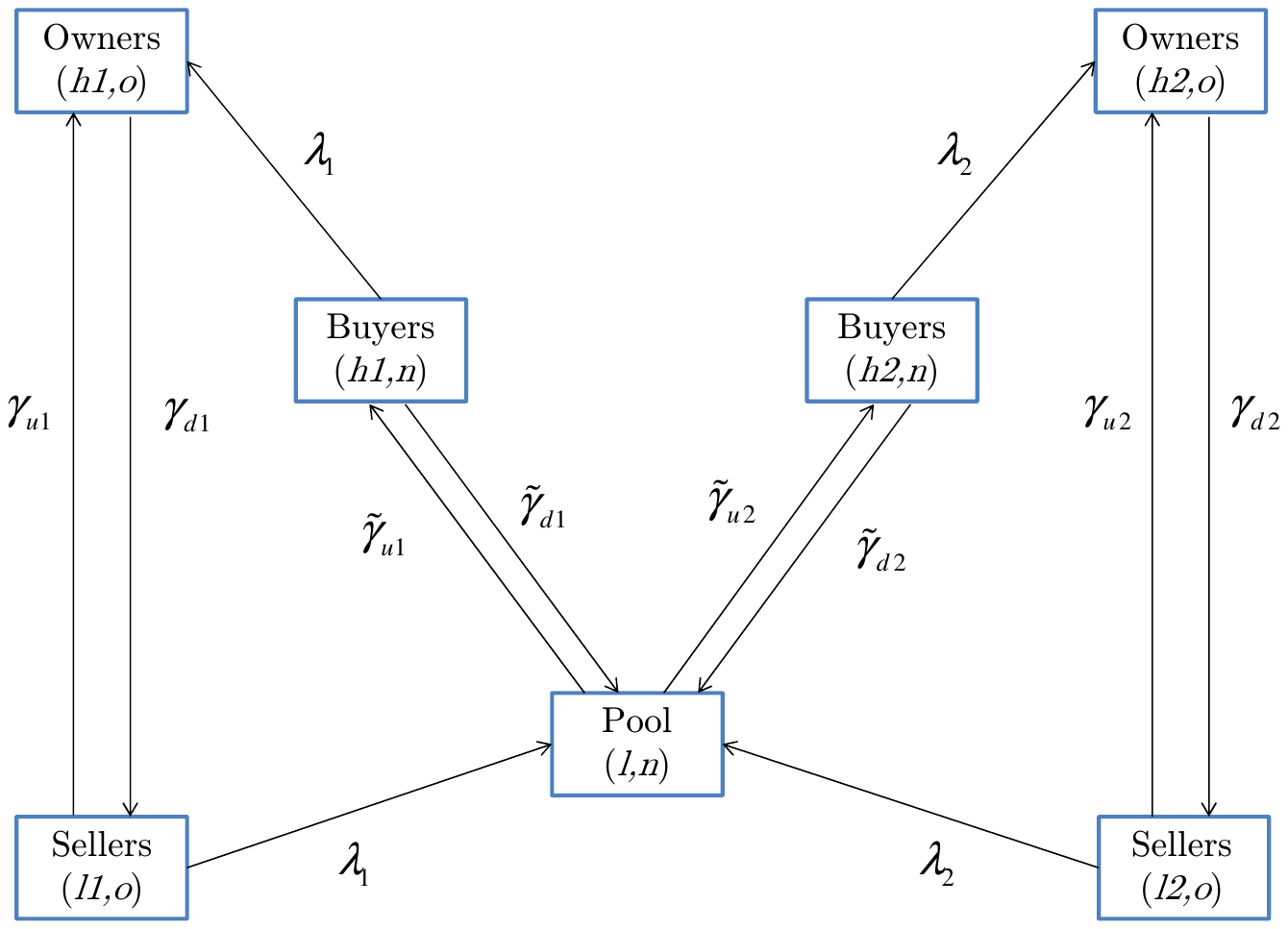}\\
  \caption{}
  \label{fig:graphSegm2assets}
\end{figure}

Since all parameters are positive, a minus sign indicates an exit from the state while a positive sign means an entrance in that state. A schematic for the dynamics between investors for this model in a two asset-market ($K=2$) is illustrated on Figure \ref{fig:graphSegm2assets}.

Note that equation (\ref{peq2}) of the previous system can be eliminated by adding each equation of (\ref{peq1}) to it. Similarly, each equation of (\ref{peq3}) can be eliminated by adding it to the corresponding equation of (\ref{peq4}). The system is then reduced to the following system of $2K$ equations:
\begin{align}\label{eq:pmasterSystem}
\begin{split}
	\dot{\mu}_t(hi,n) &= -\lambda_i \mu_t(hi,n)\mu_t(li,o) + \widetilde{\gamma}_{ui} \mu_t(l,n) - \widetilde{\gamma}_{di}\mu_t(hi,n), \ \forall i\in\mathcal{I}\\
	\dot{\mu}_t(li,o) &=  -\lambda_i \mu_t(hi,n)\mu_t(li,o) - \gamma_{ui}\mu_t(li,o) + \gamma_{di}\mu_t(hi,o), \ \forall i\in\mathcal{I}
\end{split}
\end{align}
with the $K+1$ constraints
\begin{align}
	\mu_t(hi,o) + \mu_t(li,o) &= m_i, \ \forall i\in\mathcal{I}\label{pconstraint1}\\
	\sum_{i\in\mathcal{I}} m_i + \sum_{i\in\mathcal{I}} \mu_t(hi,n)  + \mu_t(l,n) &= 1\label{pconstraint2}
\end{align}
The system (5) defines the Master Equation of our market model. It can be obtained, as in Ferland-Giroux \cite{Ferland2008}, by a functional law of large numbers or by rewriting the system with a single probability kernel, the convex combination of the two kernels, and apply Theorem 1 of B\'elanger-Giroux \cite{Belanger2013}. One could also appeal to the exact law of large numbers of Sun \cite{Sun2006} and Duffie-Sun \cite{Duffiesun2012}.

\section{Existence of a steady state for the ODE system}
The steady state is reached when all of the investors' state proportions remain constant in time, i.e. when all the derivatives appearing on the left-hand side of the above equations are zero.

From our Master Equation (\ref{eq:pmasterSystem}), we need to solve the following system of equations:
\begin{align}
	0 &=-\lambda_i \mu(hi,n)\mu(li,o) + \widetilde{\gamma}_{ui} \mu(l,n) - \widetilde{\gamma}_{di}\mu(hi,n), \ \forall i\in\mathcal{I} \label{eq:psolveEq1} \\
0 &= -\lambda_i \mu(hi,n)\mu(li,o) - \gamma_{ui}\mu(li,o) + \gamma_{di}\mu(hi,o), \ \forall i\in\mathcal{I}\label{eq:psolveEq2}
\end{align}
with the constraints
\begin{align}
	\mu(hi,o) + \mu(li,o) &= m_i, \ \forall i\in\mathcal{I} \label{eq:psolveConstr1} \\
	\sum_{i\in\mathcal{I}} m_i + \sum_{i\in\mathcal{I}} \mu(hi,n)  + \mu(l,n) &= 1 \label{eq:psolveConstr2}
\end{align}
Using each of the constraint (\ref{eq:psolveConstr1}) and substituting them in each equation of (\ref{eq:psolveEq2}) for $\mu(hi,o)$, we get
\begin{align*}
0 &= -\lambda_i \mu(hi,n)\mu(li,o) - \gamma_{ui}\mu(li,o) -\gamma_{di}\mu(li,o) + \gamma_{di}m_i, \ \forall i\in\mathcal{I}
\end{align*}
and thus
\begin{align}
\mu(li,o) &= \frac{\gamma_{di} m_i}{\lambda_i \mu(hi,n) + \gamma_i}, \ \forall i\in\mathcal{I} \label{eq:steadyMulio}
\end{align}
where $\gamma_i \triangleq \gamma_{ui}  + \gamma_{di}$. Let similarly $\widetilde{\gamma}_i \triangleq \widetilde{\gamma}_{ui}  + \widetilde{\gamma}_{di}$.

Now, subtracting each \eqref{eq:psolveEq2} to each \eqref{eq:psolveEq1} and using constraint \eqref{eq:psolveConstr2} to substitute for $\mu(l,n)$, we get:

\begin{align*}
0&=\widetilde{\gamma}_{ui} \left[1-\sum_{i\in\mathcal{I}} m_i-\sum_{i\in\mathcal{I}} \mu(hi,n) \right]-\widetilde{\gamma}_{di}\mu(hi,n) + \gamma_{ui}\mu(li,o)-\gamma_{di}\mu(hi,o), \ \forall i\in\mathcal{I}\end{align*}
then
\begin{align*}
 \widetilde{\gamma}_i \mu(hi,n)  &=  \widetilde{\gamma}_{ui}\left(1 - \sum_{i\in\mathcal{I}} m_i \right) - \widetilde{\gamma}_{ui}  \sum_{j\neq i} \mu(hj,n ) + \gamma_{ui}\mu(li,o) - \gamma_{di}\mu(hi,o), \ \forall i\in\mathcal{I}
\end{align*}

Using constraint \eqref{eq:psolveConstr1} to substitute for $\mu(hi,o)$ and substituting \eqref{eq:steadyMulio} for $\mu(li,o)$, we finally get:

\begin{align}
\begin{split}
\mu (hi,n) &= -\frac{\widetilde{\gamma}_{ui}}{\widetilde{\gamma}_{i}} \sum_{j\neq i} \mu(hj,n)+\frac{%
\gamma_{i}\gamma_{di}m_{i}}{\widetilde{\gamma}_{i}(\lambda_i
\mu(hi,n)+\gamma_{i})}\\
&\qquad -\frac{\gamma_{di}}{%
\widetilde{\gamma} _{i}}m_{i} + \frac{\widetilde{\gamma}_{ui}}{\widetilde{\gamma}_i}\left(1-\sum_{i\in\mathcal{I}}m_{i}\right), \ \forall i\in\mathcal{I}\end{split}\label{eq:steadyMuhin}
\end{align}

\noindent Hence, we have to solve a nonlinear system of $K$ equations in $K$ unknowns, $\mu(hi,n)$. Once we have solved for $\mu(hi,n)$, we can get $\mu(li,o)$ by \eqref{eq:steadyMulio} and deduce that $\mu(hi,o) = m_i - \mu(li,o)$, by \eqref{eq:psolveConstr1}, and that $\mu(l,n) = 1- \sum_{i\in\mathcal{I}}m_i - \sum_{i\in\mathcal{I}} \mu(hi,n)$, by \eqref{eq:psolveConstr2}.

We will now set things up so that we can appeal to the Leray-Krasnosel'skij-Schauder fixed-point theory.

Let $x=(x_{1},x_{2},...,x_{K})$ denote a vector in $\mathbb{R}^{K}$. Let  $m \triangleq \sum_{i\in\mathcal{I}} m_i.$
And let $I=[0,1]^{K}$ with $f=(f_1,f_2,...,f_K):I\longrightarrow \mathbb{R}^{K}$, defined by:

\begin{align}
\begin{split}
 f_{i}(x) &= x_{i}+\frac{\widetilde{\gamma}_{ui}}{\widetilde{\gamma}_{i}} \sum_{j\neq i} x_{j}-\frac{%
\gamma_{i}\gamma_{di}m_{i}}{\widetilde{\gamma}_{i}(\lambda_i
x_{i}+\gamma_{i})}\\
&\qquad +\frac{\gamma_{di}}{%
\widetilde{\gamma} _{i}}m_{i} - \frac{\widetilde{\gamma}_{ui}}{\widetilde{\gamma}_i}\left(1-m\right), \ \forall i\in\mathcal{I}\end{split}\label{eq:steadyMuhin}
\end{align}

Since each $f_i$, $1\leq i\leq K$, is a continuous map, $f$ is a continuous map.\\

Now let us consider the function $g(x)=x-f(x)$. Then $g(x)=(g_{1}(x),...,g_{K}(x))$ with
\begin{align}
\begin{split}
g_{i}(x) &= -\frac{\widetilde{\gamma}_{ui}}{\widetilde{\gamma}_{i}} \sum_{j\neq i} x_{j}-\frac{%
\lambda_{i}\gamma_{di}m_{i}x_{i}}{\widetilde{\gamma}_{i}(\lambda_i
x_{i}+\gamma_{i})}\\
&\qquad + \frac{\widetilde{\gamma}_{ui}}{\widetilde{\gamma}_i}(1-m), \ \forall i\in\mathcal{I}\end{split}
\end{align}
We will show that the continuous function $g$ has a fixed point $x^*$ which will imply that $f(x^*)=0$ and hence $x^*$ is the desired steady state of the system.
\medskip

We now state the Leray-Schauder-Krasnosel'skij theorem \cite{Djebali2010}. The theorem is valid in any Banach space, $X$. We are only considering finite dimensional spaces, so we have equivalence between the weak and the strong topologies and condition (A) of their theorem is therefore always valid for continuous functions in our situation. In fact, condition (A) is even true for bounded functions since bounded sets are relatively compact in finite-dimensional spaces.

Condition (A): Let $F$ be a function from $X$ to $X$ such that whenever $\left(x_{n}\right)_{n\in N}$ is weakly convergent, then the sequence $\left(F(x_{n})\right)_{n\in N}$  has a strongly convergent subsequence.

\begin{theo}
(Djebali-Sahnoun \cite{Djebali2010}) Let $C$ be a non-empty closed convex set in a Banach space $X$. Let $U$ be a non-empty open subset of $C$. Let $F:\bar{U}\rightarrow C$ be a continuous map which satisfies condition (A). If $F(\bar{U})$ is relatively weakly compact, then we have the following dichotomy:
(i) either the equation $F(x)=x$ has a solution in $\bar{U}$
(ii) or there exists an element $u \in \partial U$, the boundary of $U$, and there exists $p\in U$ such that $u=\lambda F(u) +(1-\lambda )p$ for some $\lambda \in (0,1)$.
\end{theo}

We will now define the domain and co-domain of $g$ in order to apply the above theorem and show that the second alternative for $F=g$ cannot happen.
Let $A_{0}= \max \{|g_{i}(x)| ; 1\leq i \leq K, x\in [0,1]^{K}\}$ and $A=\max \{A_{0}, 1\}$. We then consider the two bounded convex sets $C=[-A,A]^{K}$ and
$U=\{(x_{1},...,x_{K}): x_{i}>0 $ and $ \sum_{i\in\mathcal{I}}x_{i}< 1-m\}$. $U$ is the interior of the $K$-simplex $\bar{U}$ scaled by $1-m$. We also have that $\partial U=U_{0}\bigcup U_{1}$ with\\
$U_{0}=\{(x_{1},...,\check{x}_{i},...,x_{K}): $ with $x_j$  $\geq 0 $ $ \forall j \in \mathcal{I}  $ , where $ \check{x}_{i}=0 $ for some $ i\in\mathcal{I}  $ and $ \sum_{i\in \mathcal{I}} x_{i}<1-m\}$ and $U_{1}= \{(x_{1},...,x_{i},...,x_{K}): $ where $ x_{i}\geq 0 $ and $ \sum_{i \in \mathcal{I}} x_{i}=1-m\}$.

Suppose first that there is an element $x\in U_{0}$ for which there exist $p \in U$ and $\lambda \in (0,1)$ such that $x=\lambda g(x) +(1-\lambda )p$. In the $i^{th}$ coordinate, we then have $0=\lambda g_{i}(x) +(1-\lambda )p_{i}$. But
$$g_{i}(x) = -\frac{\widetilde{\gamma}_{ui}}{\widetilde{\gamma}_{i}} \left( \sum_{j\neq i} x_{j}-(1-m)\right) \geq 0.$$
Hence $p_{i}=-\frac{\lambda}{1-\lambda}g_{i}(x) \leq 0$ which cannot be if $p\in U$.

Suppose now that there  is an element $x\in U_{1}$ for which there exist $p \in U$ and $\lambda \in (0,1)$ such that $x=\lambda g(x) +(1-\lambda )p$.
The $i^{th}$ equality is then\newline $x_{i}=\lambda g_{i}(x) +(1-\lambda )p_{i}$ with
\begin{align}
g_{i}(x) &= -\frac{\widetilde{\gamma}_{ui}}{\widetilde{\gamma}_{i}} \sum_{j\neq i} x_{j}-\frac{%
\lambda_{i}\gamma_{di}m_{i}x_{i}}{\widetilde{\gamma}_{i}(\lambda_{i}
x_{i}+\gamma_{i})}
 + \frac{\widetilde{\gamma}_{ui}}{\widetilde{\gamma}_i}(1-m)\\
 &= -\frac{\widetilde{\gamma}_{ui}}{\widetilde{\gamma}_{i}} (1-m-x_{i})-\frac{%
\lambda_{i}\gamma_{di}m_{i}x_{i}}{\widetilde{\gamma}_{i}(\lambda_i
x_{i}+\gamma_{i})} + \frac{\widetilde{\gamma}_{ui}}{\widetilde{\gamma}_i}(1-m)
\end{align}

Hence

\begin{align}
x_{i} = \lambda \left( \frac{\widetilde{\gamma}_{ui}}{\widetilde{\gamma}_i}x_{i}-\frac{%
\lambda_{i}\gamma_{di}m_{i}x_{i}}{\widetilde{\gamma}_{i}(\lambda_i
x_{i}+\gamma_{i})}\right)+(1-\lambda) p_{i}
\end{align}

So
\begin{align}
p_{i} = \frac{1}{1-\lambda} \left(1- \lambda \frac{\widetilde{\gamma}_{ui}}{\widetilde{\gamma}_i}+\frac{%
\lambda_{i}\gamma_{di}m_{i}}{\widetilde{\gamma}_{i}(\lambda_i
x_{i}+\gamma_{i})}\right)x_{i}
\end{align}

The above equality implies

\begin{align}
p_{i} > \frac{1}{1-\lambda} \left(1- \lambda \frac{\widetilde{\gamma}_{ui}}{\widetilde{\gamma}_i}\right)x_{i}
\end{align}

And we get that $p_{i} > x_{i}$ for all $i$ which, in turn, implies that\newline
$\sum_{i\in\mathcal{I}}p_{i}>\sum_{i\in\mathcal{I}}x_{i}=1-m$. Which contradicts the fact that $p \in U$.\newline
In conclusion, we have shown that:
\begin{theo}
 Any partially segmented market with $K$ assets has a steady state which belongs to the $K$-simplex  $\bar{U}=\{(u_{1},...,u_{K}): u_{i}\geq 0 \: and \:\sum_{i\in\mathcal{I}u_{i}}\leq 1-m\}$.
\end{theo}


\section{Uniqueness of the steady state for partially segmented markets}
 In order to show the uniqueness of the steady state, we will show equivalently that the fixed point of the function $g$ used to obtain the existence result is in fact unique. We will show this using Kellogg's uniqueness theorem \cite{Kellogg1976}. As before, we state the general result for Banach spaces but in our finite-dimensional setting we have that bounded sets are relatively compact.

\begin{theo} (Kellogg \cite{Kellogg1976})
Let $D$ be a non-empty open convex set in a Banach space $X$. Let $F:\bar{D}\rightarrow \bar{D}$ be a compact continuous map which is continuously Fr\'{e}chet differentiable on $D$. Suppose that
(i) for each $x\in D$, 1 is not an eigenvalue of $F'(x)$, and
(ii) for each $x\in \partial D$, the boundary of $D$, $F(x)\neq x$. Then $F$ has a unique fixed point.
\end{theo}

Kellogg's original statement of the theorem requires the function  $F$ to be from $\bar{D}$  to $\bar{D}$ to ensure that there is at least one fixed point (by Schauder's theorem). But, as it is remarked by Kellogg on page 209, the uniqueness argument is valid if we have existence of at least one fixed point and if  $F$ is a function from $\bar{D}$  to $X$ as is the case for Altman's and  Rothe's fixed point theorems. We are in an entirely similar situation with the Djebali-Sahnoun theorem.

Note that condition (ii) shows that the fixed point cannot be on the boundary so it is not a degenerate state.

Again, in our $K$-dimensional setting, the $Fr$\'{e}$chet$ $derivative$ of\newline $F(x)=(F_{1}(x),F_{2}(x),...,F_{K}(x))$ is its Jacobian matrix
$\mathcal{J}(x)=\left(\frac{\partial F_{i}}{\partial x_{j}}\right)_{i,j}$ and $F$ is $continuously$ $Fr$\'{e}$chet$ $differentiable$ whenever all its partial derivatives are continuous.
More specifically, when $F=g$, we have $\frac{\partial g_{i}}{\partial x_{i}}=-\frac{\lambda_{i}\gamma_{i}\gamma_{di}m_{i}}{\widetilde{\gamma}_i(\lambda_{i} x_{i}+\gamma_{i})^{2}}$ and $\frac{\partial g_{i}}{\partial x_{j}}=-\frac{ \widetilde{\gamma}_{ui}}{ \widetilde{\gamma}_{i}} $ when $j\neq i$.

\begin{theo}
Each partially segmented market model has a unique steady state, that is, there is a unique solution in $U$ satisfying the system (12)-(14).
\end{theo}

\begin{proof}

We recall that $U$ is a bounded,  open and convex set. We know that $g$ is a continuously differentiable map, thus it is a compact continuous map which is continuously Fr\'{e}chet differentiable on $U$. We will have shown uniqueness if we verify that $g$ satisfies conditions $(i)$ and $(ii)$ of Kellogg's theorem.\\

 For condition $(i)$, we need to show that, for all $x$ in $\bar{U}$, 1 is not an eigenvalue of $g'(x)$, that is, the linear operator $Id-g'(x)$ is invertible. This, in turn , is equivalent to  the matrix $I(x) \triangleq Id-g'(x)$ having a determinant different from zero.\\

$$
I(x)=\left[
\begin{array}{ccccc}
g_{11} & \frac{\widetilde{\gamma}_{u1}}{\widetilde{\gamma}_1}   &     &\cdots& \frac{\widetilde{\gamma}_{u1}}{\widetilde{\gamma}_1} \\
\frac{\widetilde{\gamma}_{u2}}{\widetilde{\gamma}_2} & g_{22}   &  &\vdots& \\
 &\ddots&\ddots&\ddots&  \\
\vdots&  &  \frac{\widetilde{\gamma}_{u K-1}}{\widetilde{\gamma}_{_{K-1}}}  &     &   \frac{\widetilde{\gamma}_{u K-1}}{\widetilde{\gamma}_{K-1}} \\
 \frac{\widetilde{\gamma}_{u_{K}}}{\widetilde{\gamma}_K} &  \cdots&  &   \frac{\widetilde{\gamma}_{uK}}{\widetilde{\gamma}_K}  &  g_{KK}
\end{array}
\right]
$$

where $g_{ii}=1+\frac{\lambda_{i}\gamma_{i}\gamma_{di}m_{i}}{\widetilde{\gamma}_{i}(\lambda_{i}x_{i}+\gamma_{i})^{2}}$, for $1\leq i\leq K.$

Multiplying the $i^{th}$ line of $I(x)$ by $\frac{\widetilde{\gamma}_{i}}{\widetilde{\gamma}_{ui}}$ for each $i\in \mathcal{I}$, we get a matrix we call $J(x)$ which is equivalent to $I(x)$. More specifically,

$$J(x)=\left[
\begin{array}{ccccc}
l_{11} & 1   &     &\cdots&  1 \\
1 & l_{22}   &  &\vdots& \\
 &\ddots&\ddots&\ddots&  \\
\vdots&  &  1  &     &   1 \\
 1 &  \cdots&  &   1  &  l_{KK}
\end{array}
\right]
$$
where $l_{ii}=\frac{\widetilde{\gamma}_{i}}{\widetilde{\gamma}_{ui}} g_{ii}$, with $1\leq i\leq K$. We note for later use that $ l_{ii}>1$.

We then replace each column $C_{i}$ of $J(x)$  by the new column $C_{i}-C_{i+1}$, with
$1\leq i\leq K-1$ and obtain the equivalent matrix $H(x)$

$$H(x)=\left[
\begin{array}{ccccc}
l_{11}-1 & 0   &     &\cdots& 1 \\
1-l_{22} & l_{22}-1   &  &\vdots& \\
 &\ddots&\ddots&\ddots&  \\
\vdots&  &  1- l_{_{K-1K-1}} &     &   1 \\
 0 &  \cdots&  &   1-l_{KK}  &  l_{KK}
\end{array}
\right]
$$

The matrix $H(x)=(h_{ij})$ is
 such that $h_{iK}=1$, $h_{ii} = l_{ii}-1$ for $1\leq i\leq K-1$; $h_{KK}=l_{KK}$, $h_{ij} = 1-l_{ii}$ for $i=j+1$ and $h_{ij} = 0$ otherwise.\newline

We will  now perform elementary operations sequentially on the lines $L_{i}$ of $H(x)$ as follows:   $L'_{1}=-\frac{L_{1}}{l_{11}-1}$, and $L'_{j}=L'_{j-1}-\frac{L_{j}}{l_{jj}-1}$ for $2\leq j \leq K$. These operations give us the following upper triangular matrix

$$T(x)=\left[
\begin{array}{ccccc}
-1 & 0   &     &\cdots& t_{1K} \\
0 & -1   &  &\vdots& \\
 &\ddots&\ddots&\ddots&  \\
\vdots&  &  0 &     &   t_{K-1K} \\
 0 &  \cdots&  &   0  &  t_{KK}
\end{array}
\right]
$$

The matrix $T(x)=(t_{ij})$ is
such that $t_{ii}=-1$ for $1\leq i\leq K-1$;\\
 $t_{jK}=-\sum\limits_{\substack{i=1}}^{j}{\frac{1}{l_{ii}-1}}$, $1\leq i\leq K$
 and $t_{ij} = 0$ otherwise.\\
Its determinant  is therefore given by
$\det(T(x))=(-1)^{K}\sum\limits_{\substack{i=1}}^{K}{\frac{1}{l_{ii}-1}}$ which is different from zero. Hence 1 is not an eigenvalue of $g'(x).$\\

In order to check condition $(ii)$, we need to show that, for all $x$ in $\partial \bar{U}$ we have $g(x)\neq x.$ \newline
Assume that there exists an element $x$ in $\partial \bar{U}$ satisfying the equation $g(x)= x.$ \newline
Because $x$ belongs to $\partial \bar{U}$, then there is $i$ such that $x_{i}
=0$ or $\sum\limits_{\substack{i=1}}^{K}{x_{i}}=1-m$.
But the first case falls under the second because
$g_{i}(x) = -\frac{\widetilde{\gamma}_{ui}}{\widetilde{\gamma}_{i}} \left( \sum_{j\neq i} x_{j}-(1-m)\right) $.
So, $g_{i}(x)=0$ implies
$ \sum_{j\in \mathcal{I}} x_{j}= \sum_{j\neq i} x_{j}=(1-m)$.

\medskip
      Now for $x$ such that $ \sum_{j\in \mathcal{I}} x_{j}=(1-m)$, we recall that \newline
$g_{i}(x)= \left( \frac{\widetilde{\gamma}_{ui}}{\widetilde{\gamma}_i}x_{i}-\frac{%
\lambda_{i}\gamma_{di}m_{i}x_{i}}{\widetilde{\gamma}_{i}(\lambda_i
x_{i}+\gamma_{i})}\right).$

If $g(x)=x$ then $g_{i}(x)=x_{i}$ for all $i \in \mathcal{I}$. But this implies that either $x_{i}=0$ or
$\widetilde{\gamma}_{ui}(\lambda_{i}x_{i}+\gamma_{i})-\widetilde{\gamma}_{i}(\lambda_{i}x_{i}+\gamma_{i})-\lambda_{i}\gamma_{di}m_{i}=0$.
The latter means that $x_{i}=-\left(\frac{\widetilde{\gamma}_{di}\gamma_{i}+\lambda_{i}\gamma_{di}m_{i}}{\widetilde{\gamma}_{di}\lambda_{i}}\right)<0$.
But both cases are impossible since $x_{i}\geq 0$ and $ \sum_{j\in \mathcal{I}} x_{j}=(1-m)$.
This gives us that $g$ also satisfies condition (ii) of Kellogg's theorem and  we get the conclusion that $g$ has a unique fixed-point $x^{*}.$ This fixed point is therefore the unique point in $\bar{U}$ such that  $f(x^{*})=0$.

Thus each partially segmented market has a unique steady state.

\end{proof}

 \bigskip

\noindent \textbf{Note 1:} The matrix $H(x)=(h_{ij})$ which appears in the proof of the uniqueness theorem is
such that $h_{ij} = 0$ for $i > j+1$.  Matrices with this particular form are known in the literature as $upper$ $Hessenberg$ $matrices$. Though these matrices appear in papers such as \cite{Ramirez2014} and \cite{Janji2010} and in many control theory applications, there is no known general explicit formula of their determinant. Ramirez in  \cite{Ramirez2014} and Marjic in \cite{Janji2010} compute the determinants of  tri-diagonal Hessenberg matrices (that is the upper and lower Hessenberg matrices) and express them in terms of Fibonacci polynomials. Our proof also provides an explicit formula of the determinant in our situation.

\bigskip

\noindent \textbf{Note 2:} It is also possible to use the Poincar\'e-Miranda theorem, as formulated by Kulpa \cite{Kulpa1997}, on the hypercube $[0,r]^{K}$ for the existence result when we have the condition
$\frac{\widetilde{\gamma}_{ui}(1-m)}{\widetilde{\gamma}_{i}}\leq r\leq \frac{1-m}{K-1}$ for all $i$ in $\mathcal{I}$. If it is the case, we then have an iterative approach to approximate the steady state.  Unfortunately, when this conditon is not verified, the assumptions of the Poincar\'e-Miranda are not valid.

\textbf{Acknowledgement:}
This research is supported in part by a team grant from Fonds de Recherche du Qu\'ebec - Nature et Technologies (FRQNT grant no. 180362).



\begin{thebibliography}{99}
 


\bibitem{Belanger2013} B\'elanger, A., Giroux, G.:
\emph{Some new results on information percolation}. Stoch. Syst. 3(9), 1-10 (2013)


\bibitem{Begm2013} B\'elanger, A. Giroux, G. and Moisan-Poisson, M.:
  \emph{Over-the-counter market models with several assets}.
  q-fin.CP, arxiv:1308.2957v1.



\bibitem{Djebali2010} Djebali, S., Sahnoun, Z.:
\emph{Nonlinear Alternatives of Schauder and Krasnosels'kij types with Applications
  to Hammerstein Integral Equation to L1 Spaces}. Jour. Diff. Equat. 249, 2061-2075 (2010)


\bibitem{Duffie2001} Duffie, D.:
\emph{Dynamic Asset Pricing Theory}. Princeto.University. Press. Edit. 3. (2001)

\bibitem{Duffie2012} Duffie, D.:
\emph{Dark Markets: Asset pricing and information transmission in over-the-counter
markets}. Princeto.University. Press. Edit. 2. (2012)

\bibitem{Duffie2005} Duffie, D., G\^arleanu, N., Pedersen, L.H.:
 \emph{Over-the-Counter
Markets}. Econometrica. 73(1), 1815-1847 (2005)

\bibitem{Duffie2007} Duffie, D., G\^arleanu, N., Pedersen, L.H.:
\emph{Valuations in Over-the-Counter
Markets}. Rev.Financia. Stud. 20, 1865-1900 (2007)

\bibitem{Duffiesun2012} Duffie, D., Sun, Y.:
\emph{The exact law of large numbers for independent
random matching}. Journ. Econ. Theo. 1105-1139 (2012)


\bibitem{Ferland2008} Ferland, R., Giroux, G.:
\emph{Law of large numbers for dynamic bargaining markets}. Journ. Appl. Proba. 45(7) 45-54 (2008)




\bibitem{Janji2010} Janji\'{c}, M.:
\emph{Hessenberg Matrices and Integer Sequences}. Journ. Integ. Sequen. 1-10 (2010)

\bibitem{Kellogg1976} Kellogg, R.B.:
\emph{Uniqueness in the Schauder fixed point Theorem}. Proc. Am. Math. Soc. 60, 207-210 (1976)

\bibitem{Kulpa1997} Kulpa, W.:
\emph{The Poincar\'e-Miranda Theorem}. The Am. Math. Month. 104(4), 545-550 (1997)




\bibitem{Ramirez2014} Ram\'{I}rez, J.L.:
\emph{On convolved generalized Fibonacci and Lucas polynomials}. Appl. Math. Computation. 229, 208-213 (2014)

\bibitem{Sun2006} Sun, Y.:
\emph{The exact law of large numbers via fubini extension and characteri-
zation of insurable risks}. Journ. Econ. Theo. 126, 31-69 (2006)



\bibitem{Vayanos2007} Vayanos, D., Wang, T.:
\emph{Search and Endogeneous Concentration of Liquidity in Asset Markets}. Journ. Econ. Theo. 166(11), 66-104 (2007)




\bibitem{Weill2008} Weill, P.O.:
\emph{Liquidity premia in dynamic bargaining markets}. Journ. Econ. Theo. 140, 66-96 (2008)



\end{thebibliography}


\end{document}